\documentclass[12pt]{article}
\usepackage[dvips]{graphicx}
\begin{document}
\title{Quasinormal modes of the electrically charged dilaton
black hole}
\author{R.A.Konoplya \\
Department of Physics, Dniepropetrovsk National University\\
St. Naukova 13, Dniepropetrovsk  49050, Ukraine\\
konoplya@ff.dsu.dp.ua}
\date{}
\maketitle
\thispagestyle{empty}
\begin{abstract}
We sketch the results of calculations of the quasinormal frequencies
of the electrically charged dilaton black hole.
At the earlier
phase of evaporation ($Q$ is less than $0.7-0.8M$),
the dilaton black hole "rings" with the complex
frequencies which differ negligibly from those of the Reissner-Nordstr\"om
black hole. The spectrum of the frequencies weakly depends upon the
dilaton coupling.
\end{abstract}

When perturbing a black hole there appear damped oscillations
with complex frequencies which are the  eigenvalues of the
wave equation satisfying the appropriate boundary conditions.
Usually these are the requirements of purely outgoing waves near
infinity and purely
ingoing near the horizon. Both the complex part of the QN frequency
(inversely proportional
to the damping time) and the real one (representing the actual frequency
of the oscillation) are independent of the initial perturbations and
thereby characterize a black hole itself.
The  quasinormal spectrum of the neutron stars and black holes
is intensively investigated now, since it
is in the suggested range of the gravitational wave detectors
(LIGO, VIRGO, GEO600, SPHERE)  which are under construction.

Frequencies of the quasinormal modes of the electrically charged BH were
calculated in several papers long time ago (see \cite{Chandrasekhar} and
references therein). Yet, on various
ground, the main of which are suggestions of supergravity,
one ascribes to a black hole a scalar (dilaton) field. The latter changes
properties of a black hole, and it seems
interesting to find out what will happen to the quasinormal spectrum
when adding a dilaton charge to a black hole. Certainly,
one should expect that
for small charges of the electromagnetic  and dilaton fields the
spectrum will not differ seemingly from  that of the R-N black hole,
and, even though the black holes we see today, apparently, do not have
large electric charge, the problem is of interest, since in
charged environment electromagnetic waves will lead to gravitational
ones thereby giving a simple model for studying of the conversing of
gravitational energy into  electromagnetic one and vice versa.

We  shall consider theories including coupling gravitational,
electromagnetic and
scalar fields with the action:
\begin{equation}\label{1}
S=\int d^{4}x \sqrt{-g}(R-2(\nabla \Phi)^{2} + e^{-2 a \Phi}
F^{2})
\end{equation}

A static spherically symmetric solution of the equations following
from this action represents, in particular, electrically charged
dilaton black hole with the metric in the form:
\begin{equation}\label{2}
ds^{2} = \lambda^{2} dt^{2} - \lambda^{-2}dr^{2} -R^{2} d\theta^{2}
-R^{2} \sin^{2}\theta d\varphi^{2}
\end{equation}
where
\begin{equation}\label{3}
\lambda^{2}=\left(1-\frac{r_{+}}{r}\right)
\left(1-\frac{r_{-}}{r}\right)^{\frac{1-a^{2}}{1+a^{2}}},\qquad
R^{2}=r^{2}\left(1-\frac{r_{-}}{r}\right)^{\frac{2a^{2}}{1+a^{2}}},
\end{equation}
and
\begin{equation}\label{4}
2 M = r_{+} +\left(\frac{1-a^{2}}{1+a^{2}}\right) r_{-},\qquad
Q^{2} = \frac{r_{-}r_{+}}{1+a^{2}}.
\end{equation}
Here the dilaton and electromagnetic fields are given by the
formulas:
\begin{equation}\label{5}
e^{2 a \Phi}
=\left(1-\frac{r_{-}}{r}\right)^{\frac{2a^{2}}{1+a^{2}}}, \qquad
F_{tr} = \frac{e^{2 a \Phi} Q}{R^{2}},
\end{equation}
where $a$ is a non-negative dimensionless value representing coupling.
The case $a =0$ corresponds to the classical
Reissner-Nordstr\"om metric, the case $a=1$ is suggested by the low
energy limit of the superstring theory.
The uniqueness of static, asymptotically flat spacetimes with
non-degenerate black holes in
Einstein-Maxwell-dilaton theory was proved recently when either
$a = 1$,
or $a$ is arbitrary  but one of the fields, electric or magnetic,
is vanishing \cite{Mars-Simon}.

The perturbations obey the wave equations:
\begin{equation}\label{6}
\left(\frac{d^{2}}{dr_{\ast}^{2}} + \sigma^{2}\right)Z_{1,2}= U_{1,2}Z_{1,2}
\end{equation}
governed in axial case by the effective potentials:
\begin{equation}\label{7}
U_{1,2} = \frac{1}{2}\left(V_{1} + V_{2} \pm
\sqrt{(V_{1} - V_{2})^{2} + 4 V^{2}_{12}}\right),
\end{equation}
where
\begin{equation}\label{8}
V_{1} = a^{2} (\Phi_{,r^{\ast}})^{2} - a \Phi_{,r^{\ast} r^{\ast}}+
(\mu^{2}+2)\lambda^{2} R^{-2} + 4Q^{2} \lambda^{2} e^{2 a \Phi}
R^{-4},
\end{equation}
\begin{equation}\label{9}
V_{2} = 2 R^{-2} (R_{,r^{\ast}})^{2} - R^{-1} R_{,r^{\ast} r^{\ast}}+
\mu^{2} \lambda^{2} R^{-2},\qquad
V_{12} = -2 Q \mu e^{a \Phi} \lambda^{2} R^{-3}.
\end{equation}
Here $ dr = \lambda^{2} dr^{\ast}$ and $\mu^{2} = l(l+1)-2$, where $l$ is
the angular harmonic index. The wave equation governed by the first (second) potential at
$Q=0$ corresponds to the electromagnetic  (gravitational)
perturbations, and at $Q\neq 0$ each QN-mode will be connected with
emission of both electromagnetic and gravitational radiation.
These potentials were obtained in the
work \cite{Wilczek}, where a complete analysis of the
perturbations of the dilaton black hole was done. We have calculated
the complex quasinormal frequencies for the above class of black
holes.

Finding of the quasinormal frequencies
for black holes with
reasonable accuracy is  not such a time consuming process any more:
one can use the third oder WKB formula $(1.5)$ of the
paper \cite{Iyer-Will}, and then, in oder to improve accuracy,
use the obtained values of the
frequencies as initial guesses in the Chandrasekhar-Detweiler numerical
method \cite{Chandrasekhar-Detweiler}.For a recent review of the
methods see \cite{Kokkotas-Schmidt}.
In the present paper we were
restricted by the lower overtone modes, as those dominating in a
signal.

First, we observed that in the axial case the complex QN-frequencies
corresponding to the gravitational perturbations almost do not depend
on the value of the coupling $a$ of the dilaton field in the
wide range from $a=0$ up to $a\sim100$, unless the electric charge
(in mass units) is too large ($Q\simeq 0.7-0.8 M$). We illustrate this for
fundamental modes, i.e. for modes with $l=1$, $n=0$, where $n$ is the
overtone number in Tab.1.
This dependence on $a$ is still weak for the electromagnetic perturbations.

\begin{center}
Tab.1 The fundamental quasi-normal frequencies corresponding
to the gravitational perturbations, axial case.
\end{center}

\begin{center}
\begin{tabular}{|c|c|c|c|c|c|}
\hline
\multicolumn{3}{|c|}{$Q=0.2$}&
\multicolumn{3}{|c|}{$Q=0.9$} \\
\hline
$a$ & $Re(\omega)$ & $-Im(\omega)$  & $a$ &  $Re(\omega)$  &  $-Im(\omega)$ \\
\hline
$0$ & $0.11252$ & $0.10040$  & $0$ & $0.13275$  & $0.09980$ \\
$2$ & $0.11252$ & $0.10040$  &  $2$ & $0.13190$  &  $0.10098$ \\
$4$ & $0.11251$ & $0.10041$  &   $4$ &   $0.12936$ & $0.10294$ \\
$8$ & $0.11249$ & $0.10043$  &    $8$ &  $0.12611$  & $0.10431$ \\
$16$ & $0.11242$ & $0.10044$  &  $16$ &   & \\
$100$ & $0.11198$ & $0.10047$ &   $100$ &   &   \\  \hline
\end{tabular}
\end{center}

Certainly, the more $a$, the less charge, at which
the discrepancy with the Reissner-Nordstr\"om QN-frequencies becomes
considerable, but at the earlier stage of evaporation
the dilaton black hole "rings"
with the frequencies which are negligibly different from the
R-N frequencies, no matter
the value of the coupling constant from the above region.

From the Fig.1-Fig.3 we see that as for the classical R-N solution,
in the $a=1$ case, which is of our main interest, the real frequencies
increase with increasing of the electric charge, and the inverse
damping times are increasing also up to some maximal value at a large charge
and then falling off.
However this picture takes place for the dilaton black hole with
some kind of "retarding": the real part
of the quasi-normal frequency is less than that of the Reissner-Nordstr\"om
black hole with the same charge, and,
the corresponding inverse damping time is greater
than that of the Reissner-Nordstr\"om. This tendency
can easily be explained if taking
into consideration that the dilaton field contributes an extra
attractive field which partly compensate the effect from the increased
electric charge, thereby inducing the above changes in the
quasinormal spectrum.

Somewhat surprisingly, the real part
of the quasinormal modes
corresponding to the gravitational perturbations of the $a=1$
dilaton black hole shows no more than
$0.6$  percent "relative deviation" from R-N for any charge
$0 < Q \leq 0.99 M$, i.e.

\begin{equation}
\mathop{\mathrm{Re}}\omega_{\mathrm{dilaton BH}} =
(1-\epsilon) \mathop{\mathrm{Re}} \omega_{\mathrm{R-N BH}},
\end{equation}\
where
\begin{equation}\
\max\epsilon \approx 0.006\mathop{\mathrm{Re}}
 \omega_{\mathrm{R-N BH}},\qquad
\epsilon >0.
\end{equation}

Nevertheless, the corresponding damping times differ
seemingly when the charge is of oder $0.7-0.8 M$.

For large $l$, from the first order WKB method \cite{Will-Schutz} we obtain
approximate formula for the slowest damped mode:
\begin{equation}\label{a}
\mathrm{Re} \omega \approx \frac{1}{r_{0}}
(r_{0}- 2 M)^{\frac{1}{2}} \left(r_{0}-\frac{Q^{2}}{M}\right)^{-\frac{1}{2}}
\left(l+\frac{1}{2}\right)
\end{equation}
$$\mathrm{Im} \omega \approx -\frac{1}{2 r_{0}^{2} (M r_{0} -Q^{2})}
(30M^{4}r_{0}^{2} +Q^{4}r_{0}^{2} - 3MQ^{2}r_{0}(3Q^{2}+r_{0}^{2})-$$
\begin{equation}\label{b}
2M^{3}(21Q^{2}r_{0} +10r_{0}^{3}) +
M^{2}(16Q^{4} +25Q^{2}r_{0}^{2} +3r_{0}^{4}))^{\frac{1}{2}}
\end{equation}
where $r_{0}$ is the value of $r$ where the black hole potential
attains its maximum;
\begin{equation}\label{c}
4 M r_{0} \approx 6M^{2} + Q^{2} + \sqrt{36M^{4} -20M^{2}Q^{2}
+Q^{4}}.
\end{equation}
It is clear that, in full analogy with the
Reissner-Nordstr\"om black hole behavior,
$\mathrm{Im} \omega$ of the $a=1$ dilaton black hole
in the
large-$l$ limit tends to a negative constant,
while $\mathrm{Re} \omega$ increases linearly with $l$. From Fig.4
one can see that at large $l$, $\omega$ shows the same behavior with
changing $Q$ as for small one.

As is known when perturbing the R-N black hole, the QN-modes induced by the
axial and polar perturbations are identical \cite{Chandrasekhar}.
In addition, the R-N QN-modes
corresponding to the gravitational and electromagnetic perturbations
coincide in the extremal limit \cite{Onozawa1},
supporting the fact that only the extremal
black hole preserves supersymmetry \cite{Onozawa2}.
Both these symmetries are broken in the case of the electrically charged
dilaton black hole.
\footnote{Even though in the $a=1$
case we could not compute the QN-frequencies accurately enough when
approaching too close to the extremal limit  with the unmodified
Chandrasekhar-Detweiler or WKB methods
due to the broadening of the effective potentials \cite{Wilczek},
the values of the quasinormal frequencies we
obtained for $Q =1.41 M$ do not leave any hope that the frequencies
for gravitational and electromagnetic perturbations
will coincide in the extremal limit.} We expect that is the
extremal dilaton black hole
with {\it both} electric and magnetic charges, being N=4 supersymmetric
when embedded in N=4 supergravity \cite{Linde-Kallosh}, which must
respond in the same manner on gravitational and electromagnetic
perturbations, and to check it is the point of our future investigation.

It worthwhile mentioning, that in the axial case,
the dilaton field itself does not suffer
from perturbations unlike the polar one. Yet the polar perturbations
of the dilaton black hole
are governed by a very cumbersome potentials, and we have not found
a better way than,
following the paper \cite{Wilczek}, to consider the spectator scalar field
propagating in the
black hole background as a qualitative model for
perturbations. Consequently, the WKB accuracy was sufficient under such
qualitative consideration.
It proved out that the above general
properties of the axial modes, are valid in the considered
case as well and, apparently, spread on polar modes.

\newpage
\begin{figure}
\begin{center}
\includegraphics{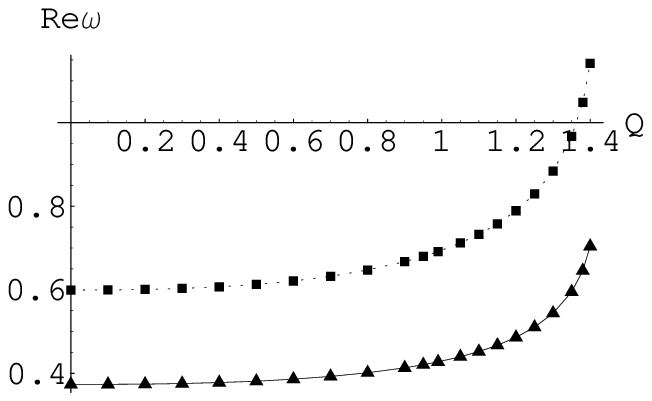} \quad \includegraphics{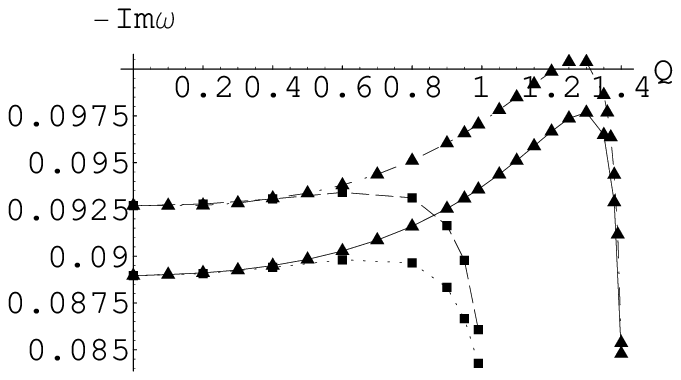}
\caption{Real and imaginary parts of $\omega$, $l=2,3$, for axial
gravitational perturbations of the
$a=1$ dilaton black hole and R-N black hole. For  $0<Q \leq 0.99 M$, $\mathrm{Re}
\omega_{\mathrm{dilaton BH}} =
(1-\epsilon) \mathop{\mathrm{Re}} \omega_{\mathrm{R-N BH}}$,
where $\max\epsilon  \approx
0.006\mathop\mathrm{Re}\mathop\omega_{\mathrm{R-N BH}}$.}
\label{gr6}
\end{center}
\end{figure}

\begin{figure}[]
\begin{center}
\includegraphics{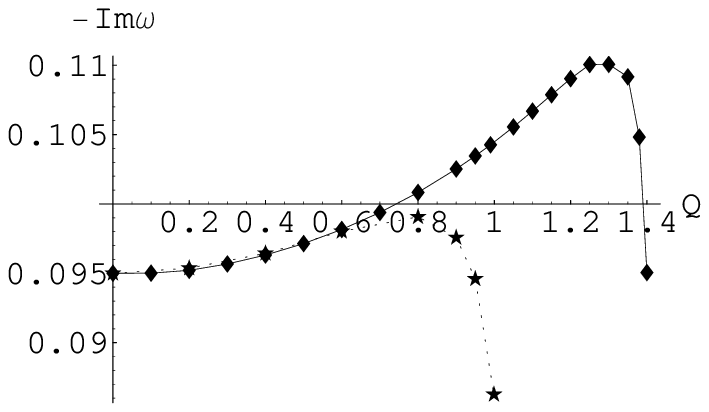} \quad \includegraphics{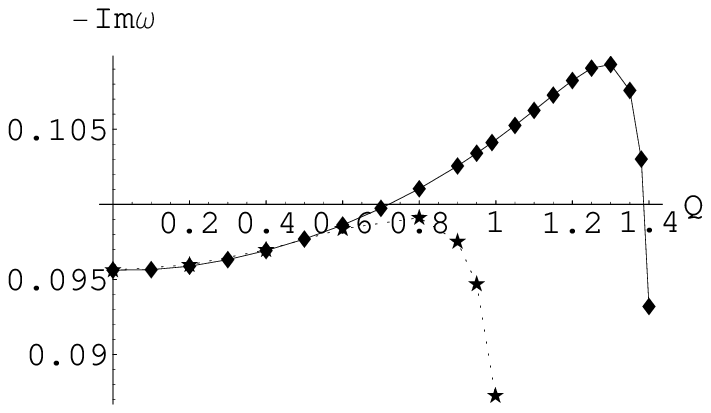}
\caption{Imaginary part of $\omega$ for  axial electro-magnetic
perturbations of the $a=1$ dilaton black hole and R-N black hole;
$l=2$ and $l=3$.}
\label{gr_4}
\end{center}
\end{figure}

\begin{figure}[]
\begin{center}
\includegraphics{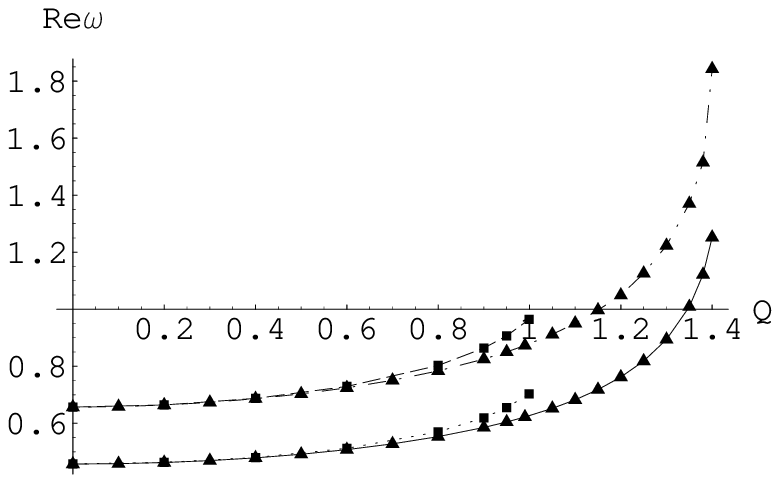} \includegraphics{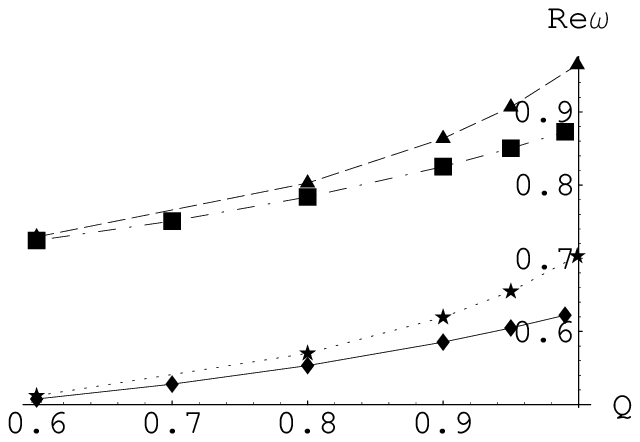}
\caption{Real part of $\omega$ for axial  electro-magnetic perturbations
($l=2,3$) for $a=1$ dilaton black hole and for R-N one.
Enlarged region of the figure
shows when the difference
between the R-N QN-modes and those of its dilaton analog
can not be ignored.}
\label{gr_5}
\end{center}
\end{figure}

\begin{figure}[]
\begin{center}
\includegraphics{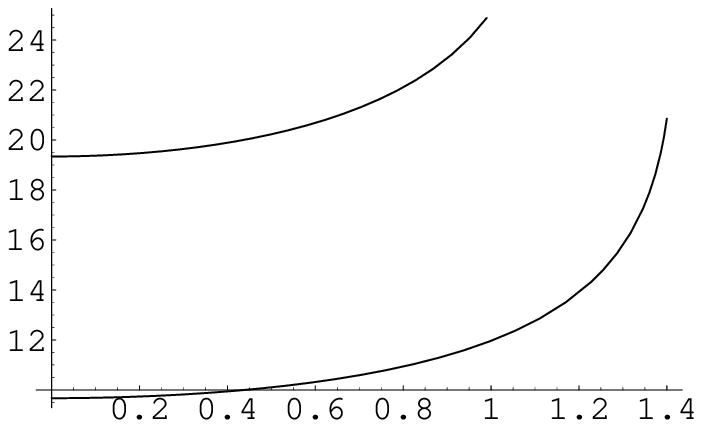} \quad \includegraphics{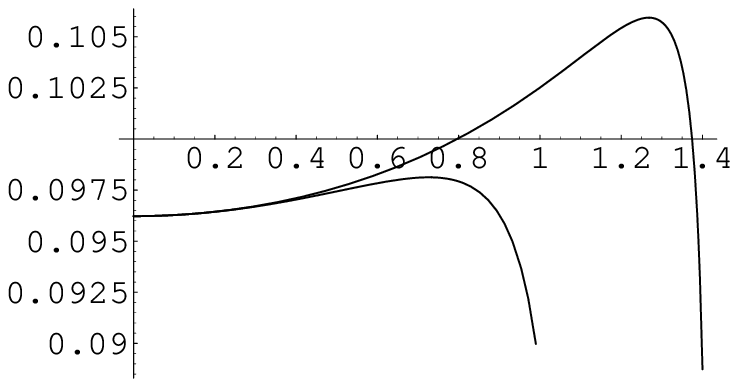}
\caption{Real and imaginary parts of  $\omega$ for large $l$ as an
approximate function of $Q$ for
$a=1$ dilaton black hole (by the formulas
(\ref{a}-\ref{c})) and for R-N (by the formulas (4-5) of the work
\cite{Andersson-Onozawa}) ($M=1$, $l=100$).}
\label{polarimaginary}
\end{center}
\end{figure}

\end{document}